# Does the short-term boost of renewable energies guarantee their stable long-term growth? Assessment of the dynamics of feed-in tariff policy

Milad Mousavian H.[1], Hamed Shakouri G.[1], Ali-Naghi Mashayekhi[3], Aliyeh Kazemi[4*]

[1] School of Industrial and Systems Engineering, College of Engineering, University of Tehran, Tehran, Iran

[3] Graduate School of Management and Economics, Sharif University of Technology, Tehran, Iran

[4] Department of Industrial Management, Faculty of Management, University of Tehran, Tehran, Iran

---

[*] Corresponding author. Tel: +982161117771; fax: +982188006477.
E-mail addresses: aliyehkazemi@ut.ac.ir
Address: Nasr Bridge, Chamran Highway, PO Box: 14155-6311, Tehran, Iran.

**Highlights:**

- An SD model is established to study the impact of FiT mechanism on REs expansion.
- The trend of REs development in Iran is analyzed for both short- and long-term horizons.
- Three scenarios are tested as policies for the development of REs.
- Social mechanisms can weaken the effect of economic incentives in long-term.
- REs tax upon the budget status is the best policy.

# Does the short-term boost of renewable energies guarantee their stable long-term growth? Assessment of the dynamics of feed-in tariff policy

**Abstract**

Feed-in tariff (FiT) is one of the most efficient ways that many governments throughout the world use to stimulate investment in renewable energies (REs) technology. For governments, financial management of the policy is very challenging as that it needs a considerable amount of budget to support RE producers during the long remuneration period. In this paper, we illuminate that the early growth of REs capacity could be a temporary boost and the system elements would backlash the policy if financial circumstances are not handled well. To show this, we chose Iran as the case, which is in the infancy period of FiT implementation. Iran started the implementation of FiT policy in 2015 aiming to achieve 5 GW of renewable capacity until 2021. Analyses show that the probable financial crisis will not only lead to inefficient REs development after the target time (2021), but may also cause the existing plants to fail. Social tolerance for paying REs tax and potential investors' trust emanated from budget-related mechanisms are taken into consideration in the system dynamics model developed in this research to reflect those financial effects, which have rarely been considered in the previous researches. To prevent the financial crisis of the FiT funding and to maintain the stable growth in long-term, three policy scenarios are analyzed: continuation of the current program with higher FiT rates, adjusting the FiT rates based on the budget status, and adjusting the tax on electricity consumption for the development of REs based on the budget status. The results demonstrate that adjusting the tax on electricity consumption for the development of REs based on budget status leads to the best policy result for a desired installed capacity development without any negative social effects and financial crises.

**Keywords**: Feed-in tariff, Renewable energies, System dynamics, Policy resistance, Social acceptance

# 1 Introduction

Finite resources and environmental degradation are two main reasons for governments thinking of providing electricity from renewable rather than non-renewable resources. By such a diversification, besides empowering energy security and retaining sustainability in production, they combat climate changes as well [1]. Renewable energies (REs) are recognized as one of the best alternatives substituting fossil fuels; nonetheless, high capital costs and changes in the level and composition of investment make them an expensive energy resources [2].

Low fossil fuel prices prevent REs to expand rapidly in the absence of effective incentives [3]. To tackle this issue, various types of policy tools including price-based incentives such as feed-in policies, quantity-based incentives or quota obligations, including renewable portfolio standards (RPS) in combination with REs certificate or credit (REC) markets, fiscal and financial incentives such as tax credits, and voluntary measures such as green tariffs have been used by governments to support REs development [2]. One of the most popular policy has been adopted by many countries is feed-in tariff (FiT). Fit is an intensive program that provides investors with a set of payments for the electricity which is produced by REs and fed into the power grid. Small-scale developers like homeowners and medium to large-scale companies can benefit from the supporting program to encourage their participation in such programs by securing definite returns of their investments [4]. When the private independent producers receive a long-term, minimum guaranteed price for the renewable electricity they generated, a certain degree of financial reliability is provided which resulted in less investment risk and more willingness to invest. This is the considerable benefit of FiT. Even though FiT is one of the most effective REs policy mechanisms in promoting and sustaining REs growth [5], it may lead to some drawbacks if it is not applied correctly. There exist some real-world examples of governments with electricity consumers facing financial burdens imposed by the FiT policy [5–7]. FiT prices, digression rates and the period in which FiT policy is applied are the most critical factors when utilizing this policy. FiT rates must be high enough to recover the investment cost within a reasonable timespan and simultaneously small enough to avoid enforcing a significant financial burden [8]. A long-term, stable and high price can negatively affect the actual energy market. When the FiT price is too high, the pace of REs growth may exceed the goal predicted by policy makers [9] which may restrict them under different economic conditions and adversely affect the investors' confidence in this incentive program [1].

The objective of this study is to diagnose the FiT policy structure and evaluate its effect on REs growth trend in long-term. A system dynamics (SD) approach is used to show the dynamic interaction of FiT policy and other factors such as potential investors' trust and social acceptance, and to test the alternative or corrective policies. Using this model, policymakers can analyze and forecast the future situation of REs regarding different policies they are considering. They can conclude the suitable amount of FiT rate to be paid for renewable electricity in different periods, and comprehend the reason for probable malfunction of the system.

For our analysis, we selected the case of Iran. Even though Iran is an energy-rich country, both energy security and contribution to fewer carbon emissions for the country require the faster development of REs. Due to the little share of REs in the current energy portfolio, expanding the electricity production from renewable resources is significantly essential [10]. However, the dynamic mechanism of the FiT system, which considers social and economic interactions in the long-term, has been rarely studied; what this research focuses on.

The structure of this present paper is as follows: Section 2 briefly reviews the relevant literature concerning the FiT and REs development. The status of REs and FiT in Iran is described in Section 3. In the next section, a brief explanation of the research methodology and the modeling process is given, and the suitability of the SD approach for investigating the problem is discussed. Section 5 presents the qualitative and quantitative descriptions of the REs development model. Section 6 discusses the simulation results considering different policies. Finally, Section 7 concludes the paper.

## 2 Literature review

FiT has appeared as one of the most popular policies for supporting renewable technologies. Several papers have discussed the advantages or disadvantages of different FiT policies as well as the potential financial difficulties created by this policy [5,7,11–14]. To evaluate the FiT policies, some researchers developed different assessment models and approaches. For instance, Dusonchet and Telaretti [15] performed an economic analysis to investigate the effect of FiT on promoting photovoltaic (PV) technology in the eastern European Union (EU) countries. The analysis showed that, in some cases, supporting policies could be inappropriate for the owner of the PV system. Also, in many cases, the difference between the implementation of the same supporting policy lead to significantly different results in different countries. Erturk [16] examined the onshore wind energy potential of Turkey to discover if FiT would help this potential. In this study, the economic analyses were conducted by the construction of a static model accompanying an uncertainty analysis in order to found out which kinds of onshore wind projects were feasible and more attractive. Bakhshi and Sadeh [17] suggested a dynamic FiT strategy can be implemented in the developing countries like Iran, where high technology equipment is imported, and economic situation is not stable. In the proposed scheme FiT was updated once a year respecting two main parameters Euro exchange rate, and good retail prices. After economic analysis and calculating indexes such as net present value (NPV), and internal rate of return (IRR), they concluded that by applying this scheme, the PV viability for short- and mid-term was guaranteed. Tabatabaei et al. [2] discussed the economic, welfare and environmental impacts of FiT policy in Iran. They examined the effect of FiT policy under different scenarios to increase the production of electrical energy from renewable resources as much as 10%. The results showed that the application of subsidies to REs and the way the government finances these subsidies could affect the results of FiT policy.

Different SD simulation models were applied successfully to a variety of problems related to FiT. In the following, we benefited from reviewing the SD models designed, developed and put into action for real-world cases.

Using the methodology of SD, Baur and Uriona [18] developed a model of the German PV market for small pants on private houses and tested public policies. Different scenarios respecting the reduction or even elimination of the FiT scheme were analyzed. They concluded public policy has a key role in the path of transition to RE patterns and consequently it has to be cautiously employed. Zhang et al. [19–22]

developed an SD model to evaluate the effect of FiT and renewable portfolio standards (RPS) on the development of China's biomass, wind, and PV power industries. The results showed that in the purely competitive market, RPS could promote PV, waste incineration, and biomass development better than the FiT; however, the integrated implementation of FiT and RPS can result in better outcomes for the wind power industry. Ye et al. [23] examined the FiT policy for PV in China. The economic tools of NPV, IRR, learning curve and the SD method were applied to analyze the dynamic mechanism of the FiT system. The finding of the study indicated that the authority should adopt the FiT more frequently, at least once every year. An SD model was designed by Hsu and Ho [13] to assess the FiT policy on wind power installation in Taiwan. They concluded that the FiT policy could lead to a reduction in greenhouse gas (GHG) emissions and development of wind power industry. Li et al. [24] discussed the paper and put forward suggestions to perfect the historical test. Castaneda et al. [25] presented an SD model to evaluate the effects of FiT policy in the British electricity market. Results suggested that FiT schema is a suitable policy tool for reaching emission reduction to a lower cost. An SD model was proposed by Ahmad et al. [1] for analyzing the role of FiT policy to promote PV investments in Malaysia. The results demonstrated that higher FiT rates resulted in higher installed PV capacity. Shahmohammadi et al. [8] propounded an SD model to evaluate the effect of the FiT mechanism on Malaysia's electricity generation mix. They concluded that albeit the policy can lead in satisfactory results, the government may encounter an increasing budget shortage and it is necessary to increase its income sources. Akhwanzada and Tahar [26] developed an SD model and analyzed the effect of FiT policy and reserve margin on the expansion of PV and municipal solid waste capacities in Malaysia. Using an SD model, Hsu [9] assessed the effects of Fit and capital subsidies on PV installations in Taiwan. They illuminated appropriate policies such as reasonable FiT prices or subsidies, and mandatory regulations can result in PV capacity development. Lyu et al. [27] created an SD model to study the influence of FiT and RPS on the installed capacity of PV and emission reduction in China. The best solution was the combination of FiT and RPS policies. Hoppmann et al. [28] analyzed the evolution of the FiT system for PV development in Germany. By investigation dynamics of the system, they explained how the characteristic of socio-technical systems affects policy interventions.

In almost all of the previous works, it is given that the government could cover the policy expenses and there would be no financial burden. While the budget and monetary mechanisms have a pivotal role in the FiT policy success, in many of the past researches, the budget mechanisms are not modeled, and only the cost of the policy or the cost of the GHGs reduction is calculated. If the mechanisms are not well-designed, then the REs development pathway could be affected adversely. This may be the root of many long-term harmful social effects on the system, the point that we focus on in this research.

## 3 The case of Iran: status of REs and FiT

REs hold a tiny share of energy production in Iran. Low fossil fuel prices and the subsidies on energy consumption are the main reasons for the low share [2]. Based on the statistical reports published by Iran's Ministry of Energy [29], the share of fossil fuels in the total primary energy supply was 98.77% in the year 2016, and the number for REs and nuclear energy were 0.94% and 0.29%, respectively. Iran's energy economy indexes reflect a high rate of energy consumption per capita. The high consumption of fossil fuels is one of the main causes of air pollution in Iran which imposes high environmental and economic costs. Four of the top ten air polled cities in the world are situated in Iran. Power supply during peak hours in summer afternoons is also a serious problem. Thus the construction of new power stations, especially renewable systems with the natural peak shaving in hot climates is compulsory [17].

There is an enormous potential for electricity production from renewable resources in Iran. Iran has appropriate areas for solar energy deployment. The annual average of solar radiation and sunny hours during different seasons has provided high potential for solar power generation in the country. Moreover, due to strong winds in several locations, more development of wind power capacity is possible. Iran also has many rivers with ideal conditions to expand hydropower plants. The potential for power production from biomass resources is also high [30]. Furthermore, on account of the fact that Iran is located on the geothermal belt, there exists a high potential for geothermal energy production.

Based on the mentioned facts, Iran's Ministry of Energy introduced new rules to promote the investment of renewable technologies. After unsuccessful net-metering and capital subsidies program during 2013-2014, the new FiT program was introduced in 2015 to convince investors to invest in renewable systems. It should be noted that that the target capacity was determined to be 5 GW until 2021. According to the new rule, all individuals, including house owners and commercial investors can produce electricity from RE systems and sell it for up to 20 years at a guaranteed price, regardless of their domestic consumption [17]. The renewable organization of Iran (SUNA) is responsible for making appropriate arrangements for the implementation of the mentioned policy.

## 4 Research methodology

This study uses the SD method to construct a behavioral model and diagnose the FiT policy structure in Iran. SD is an approach to understand the behavior of a complex system through its components. With a social system-related management concept developed by Jay W. Forrester, SD deals with interconnections, nonlinearities and complexity of systems. Causality is a basis for this approach and causal feedback loops can be realized and analyzed through systems thinking. Using computer simulations, the real influence of a policy on a social system and its consequences can be studied to understand the implied causal feedback in the system [13].

In this paper, by reviewing the related literature and studying the REs status and FiT history in Iran, the problem is articulated. Then the boundary of the model, endogenous and exogenous variables, and the corresponding relationships are determined. In the next step, a conceptual framework is formulated in which subsystems and balancing and reinforcing causal mechanisms are presented through subsystem and causal loop diagrams respectively. Next, a mathematical model is developed to simulate the current and future trends of FiT policy. Before system simulation, the validation of the model is tested. In this step, both the structural and behavior validities are examined. Finally, the current FiT policy is simulated and analyzed. Three policies consisting of continuation of the current program with higher

FiT price, adjusting the FiT price based on the budget status, and adjusting the tax on electricity consumption for renewables development (REs tax) upon the budget status, are considered, and the behavior of the system is simulated and analyzed for up to 2035.

## 5 SD model of FiT

### 5.1 Conceptual framework

Fig. 1 shows the subsystems of the model and their interactions. This diagram gives a comprehensive view of the model structure so that while paying attention to less detail, it is possible to get a better understanding of the systematic endogenous perspective of the model. There are three main sectors: budget, REs development, and FiT payment. Budget and REs development subsystems interact with each other directly and indirectly through the FiT payment subsystem. The budget subsystem, which is the source of FiT policy causes self-reinforcing mechanisms in the REs development subsystem and leads to the capacity growth of renewables. The first reinforcing loop in the REs development subsystem represents the growth of renewables' penetration rate. Where, social acceptance, and as a result, the tendency to invest in this type of electricity generation technology lead to a more installed capacity of REs. There is another reinforcing loop in this subsystem that captures the learning process effect. More installed renewable capacity leads in more experience, learning effect and less capital cost of installation. Therefore, the tendency for REs investment rises. In the budget subsystem, when the government buys the electricity generated by renewables, the budget drops, and if it cannot cover the payment, the REs tax compensates.

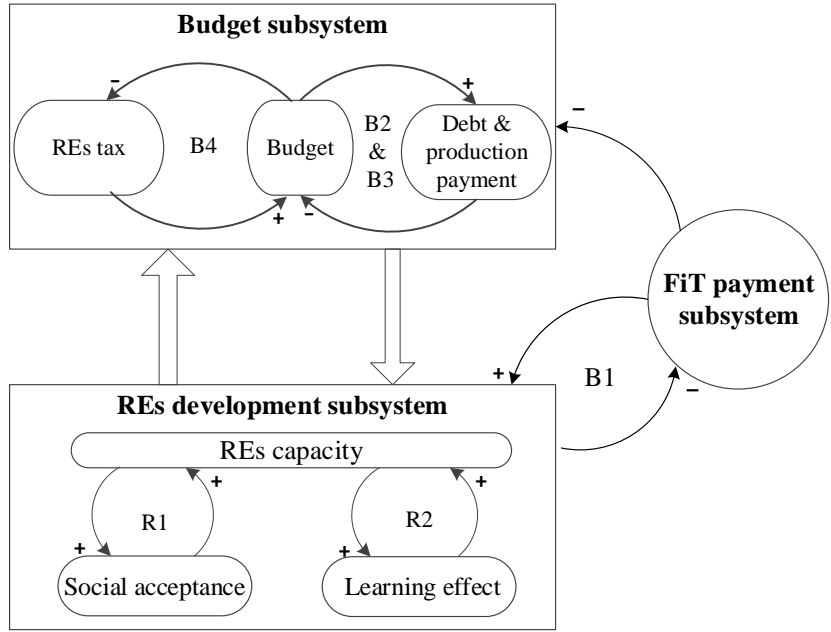

Fig. 1. Subsystem diagram of the model.

## 5.2 Causal loop diagram

The causal loop diagram for the study is shown in Fig. 2. A total of six feedback loops were identified and labeled as R1, R2, B1, B2, B3, and B4. The first two loops are reinforcing, whereas the last four are controlling or balancing loops. The reinforcing loops have an intensification effect, while the balancing loops have a limiting effect on the system. The interaction between these two types of loops generate the dynamics of the system.

Wuestenhagen et al. [31] emphasize that social acceptance is a crucial factor affecting the REs development plan implementation. They conceptualize one of the essential aspects of social acceptance by defining market acceptance, which implies the diffusion of the innovation process. We use the adoption of new consumers' concept to capture the effect of renewable diffusion and social acceptance in our model. The part of the society who may invest in the renewable projects is named as "potential investors". When the tendency to invest increases, FiT requests increases, which, in turn, leads to investment if approved by the decision makers. The higher the investment, the more the installed plants. Increasing the installed capacity leads to increasing the diffusion of renewables, which is conceptualized with the variable renewables' penetration rate in the model. More penetration rate, in turn, causes more social acceptance, higher tendency to invest, more FiT requests, and therefore, higher potential investors to actual investors. This phenomenon forms the positive feedback loop R1, namely "social acceptance".

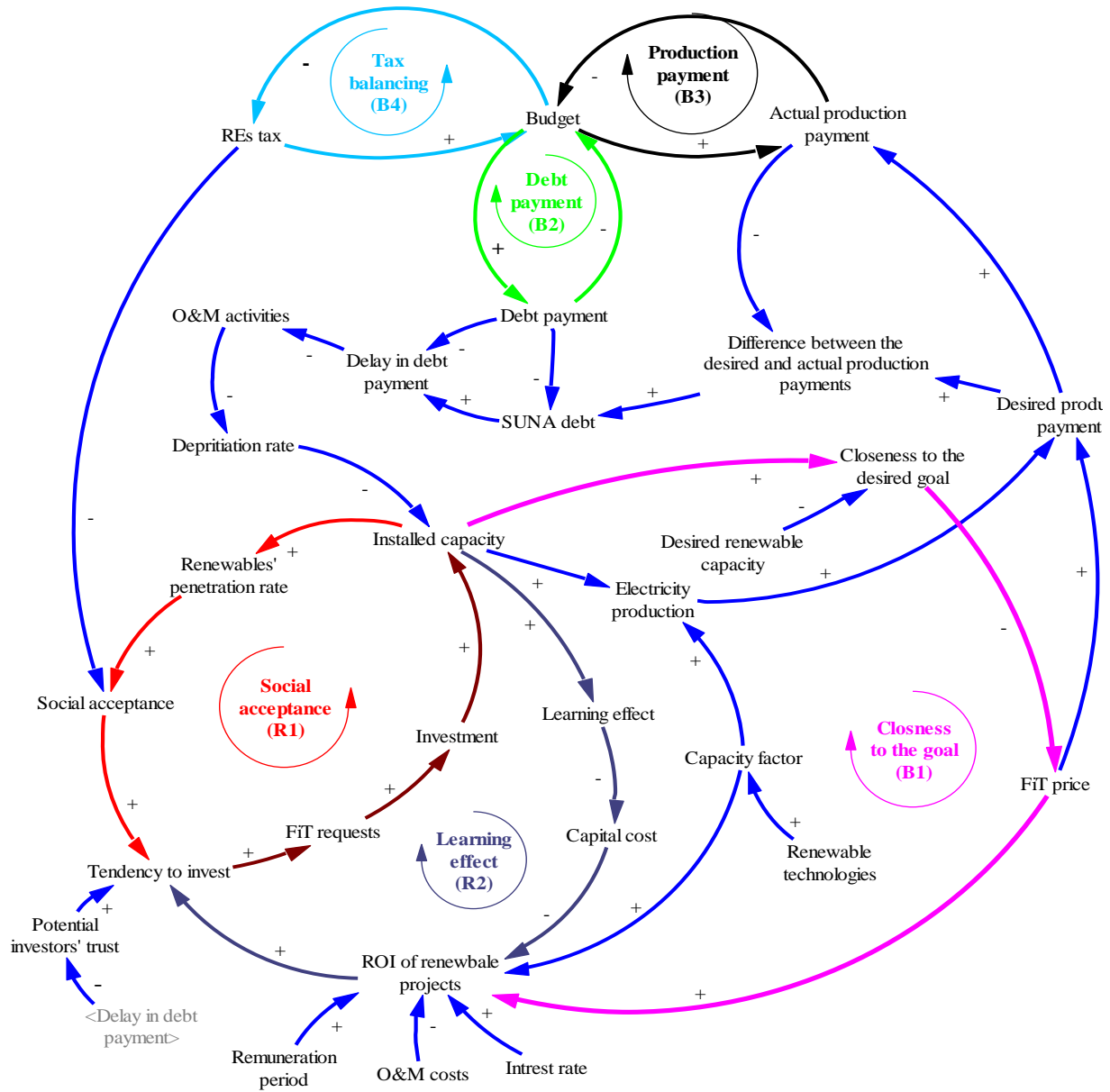

Fig. 2. Causal loop representation of FiT assessment model.

The REs capacity growth influences the experience for using and making renewable systems. This learning results in lowering the capital cost, meaning the higher return of investment (ROI) and more tendency to invest in renewable resources. It also leads to more FiT requests, higher investments and then more installed capacity. This is how the second positive feedback loop (R2) works. Learning effect is the name of this loop, which is modeled based on the research done by Hsu [32].

The gap between the government target and existing renewable capacity, which affects the FiT mechanism is one of the prevalent concepts used by some researchers like Ahmad et al. [1], Mousavian et al. [10], and Hsu [9]. When the installed capacity grows, the distance to the desired goal (5 GW installed capacity in 2021) decreases, resulting in lowering the FiT rate by the government. It causes a reduction in the ROI of renewable projects and thereby less tendency to invest, fewer request for FiT, less investment, and consequently, fewer installed capacity. This phenomenon forms the negative feedback loop B1, namely "closeness to the goal".

When the amount of budget falls below the amount of the whole desired payment to the whole desired payment to the renewable producers, a budget shortage occurs, thereby the government as the buyer is not capable of paying for the production and for the previous debts; this ultimately leads to less debt payment. Moreover, the amount that the government pays decreases the available funds, learning to more budget shortage, and thus, forming the negative feedback loop B2, namely "debt payment".

The B3 loop consists of similar variables to those of loop B2. Instead of the debt payment affecting the budget, another variable (namely "payment for electricity production") affects the amount of budget. This loop is labeled as the "production payment".

When the budget shortage is perceived by the government, it is decided to increase the REs tac whit the aim of compensating the budget shortage. It results in more amount of budget. This phenomenon forms the negative feedback loop B4, namely "REs tax balancing". However, though it is claimed that this controlling mechanism exists in the current system, the REs tax has remained constant in recent years and does not react to the budget variations. Therefore, it seems that the feedback link from budget to REs tax has not been activated so far, although according to the policy makers, it potentially exists.

## 5.3 Model stock-flow structure

Below are details of the model from the perspective of stock and flow variables, where the important mathematical equations of each subsystem are described. Stocks are accumulations and so characterized the state of the system. By decoupling the inflows and outflows and causing delays, the sources of disequilibrium dynamics in a system are specified. Vensim, an SD simulation software (Vensim PLE for Windows Version 6.0b), is going to be used to simulate the behavior of renewable installed capacity and other related mechanisms of the system for the years 2015-2035 by its six state (stock) variables. The whole stock-flow diagram is illustrated in Fig. 3, which shows the detailed relationships between the variables of the SD model with stocks, flows, auxiliary variables and parameters.

### 5.3.1 REs development

Fig. 4 shows the stock-flow diagram of installed capacity. In the simulation system, installed capacity is defined as the accumulation of construction rate minus depreciation (see row 1 in Table 1). Approved FiT requests divided by the time needed to build a renewable power plant makes the in-flow of installed capacity, namely "construction rate" (see row 2 in Table 1). Since some requests are rejected by SUNA due to the legal or qualification reasons (approximately half of annual FiT requests leads to capacity construction), a number of 0.5 is considered as the fraction of rejected requests (see row 3 in Table 1). While depreciation is an out-flow of the installed capacity, it is the in-flow of the depreciated capacity stock variable and equal to the installed capacity divided by the equipment's lifetime (see row 4 in Table 1). Cumulative installed capacity is equal to the sum of installed capacity and depreciated capacity, which is demonstrated by row 5 in Table 1. The initial value of installed capacity is set as 120 MW according to the SUNA data in 2015. The initial value of depreciated capacity equals to zero at the beginning of the simulation.

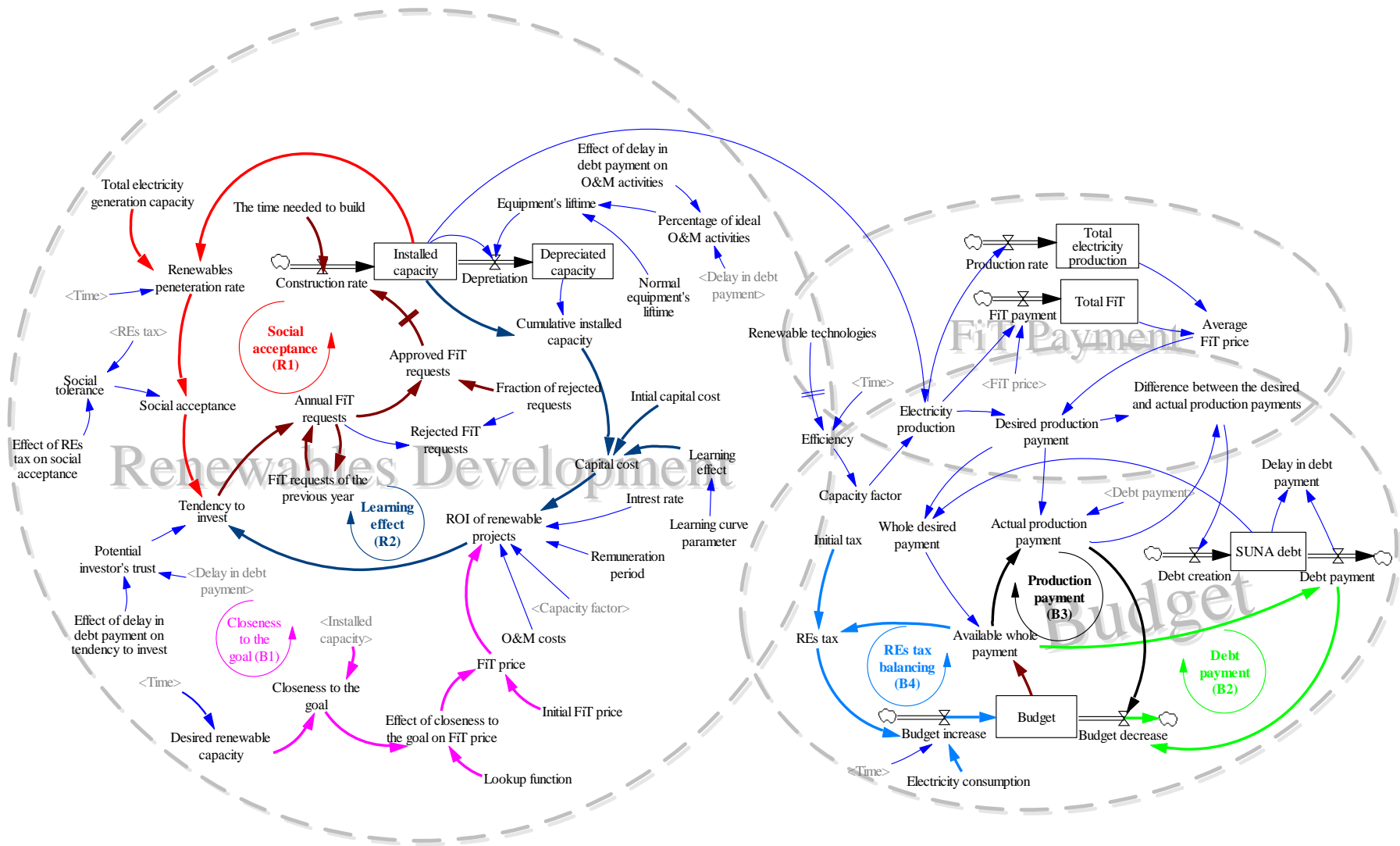

Fig. 3. Stock-flow diagram of FiT effects on REs development.

Table 1. Renewables development subsystems' mathematical equations.

| Item | Variable (Unit) | Mathematical equation |
|---|---|---|
| 1 | Installed capacity (Megawatt (MW)) | = INTEG (Construction rate - Depreciation, 120) |
| 2 | Construction rate (MW/year) | = Approved FiT requests/The time needed to build |
| 3 | Approved FiT requests (MW) | = Annual FiT requests * (1 – Fraction of rejected requests) |
| 4 | Depreciation (MW/year) | = Installed capacity/Equipment's lifetime |
| 5 | Cumulative installed capacity (MW) | = Depreciated capacity + Installed capacity |
| 6 | Annual requests for FiT (MW) | = FiT requests of the previous year * Tendency to invest |
| 7 | Tendency to invest (Dimensionless) | = ROI of renewable projects * Social acceptance * Potential investors' trust |
| 8 | Renewables' penetration rate (Dimensionless) | = Installed capacity/Total electricity generation capacity (Time-based linear regression) |
| 9 | ROI of renewable projects (Dimensionless) | = (((Capacity factor * 8760 * (FiT price - operation and maintenance (O&M) costs)) * (((1+interest rate) ^ Remuneration period - 1)/Interest rate * (1 + Interest rate) ^ Remuneration period) - Capital cost))/Capital cost |

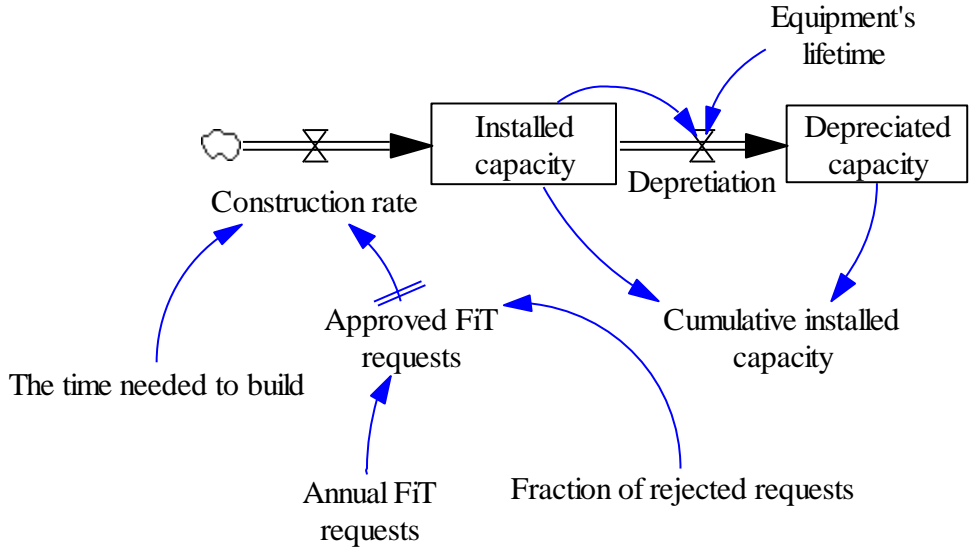

Fig. 4. Stock-flow diagram of installed capacity.

Fig. 5 demonstrates the annual FiT requests' causal relations. It is equal to the FiT requests of the previous year multiplied by the tendency to invest (see row 6 in Table 1). In this study, it is assumed that the public tendency to invest for REs is correlated with the ROI of renewable projects, social acceptance, and the potential investors' trust (see row 7 in Table 1). Renewables' penetration rate, which is one of the factors affecting social acceptance, is equal to the installed capacity of REs divided by the total electricity generation capacity, which is calculated through a time-based linear regression of historical data (see row 8 in Table 1).

Fig. 6 displays the causal relations of ROI of renewable projects. The decision about investment in REs projects is based on their ROI that is a performance measure, which is used to evaluate the efficiency of an investment. ROI measures the amount of return of investment, relative to the investment's cost. To calculate ROI of renewable projects, the benefit (or return) of the investment is divided by the cost of the investment (see row 9 in Table 1). The remuneration period refers to the time horizon that SUNA is obliged to purchase the electricity produced by REs and fed into the grid. According to the SUNA regulations, this period is 20 years [33].

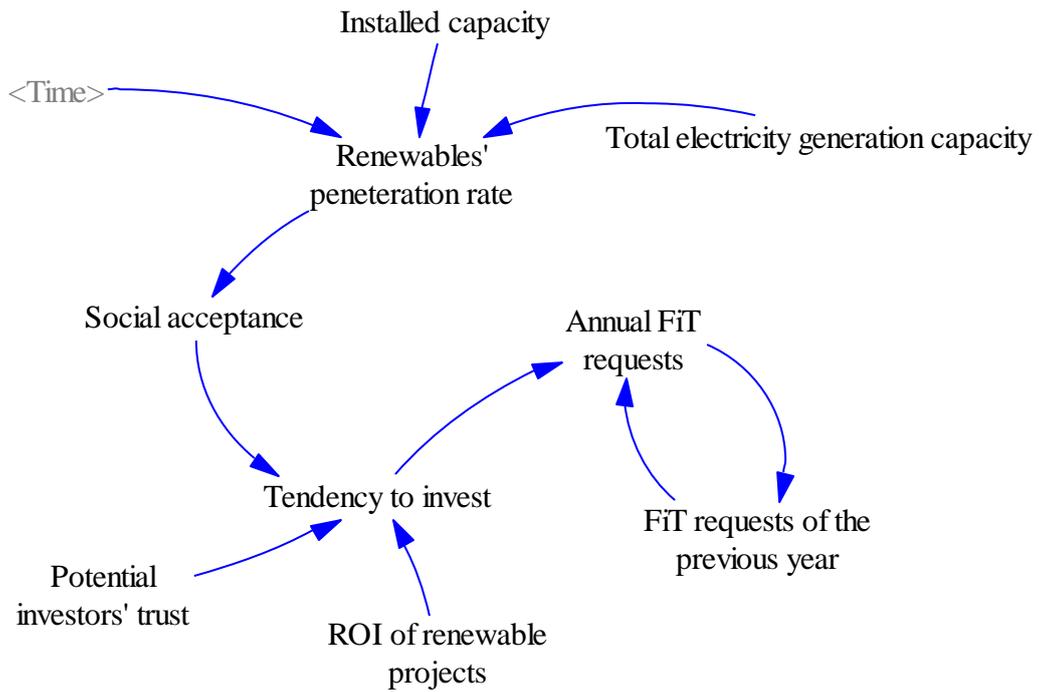

Fig. 5. Annual FiT requests' causal relations.

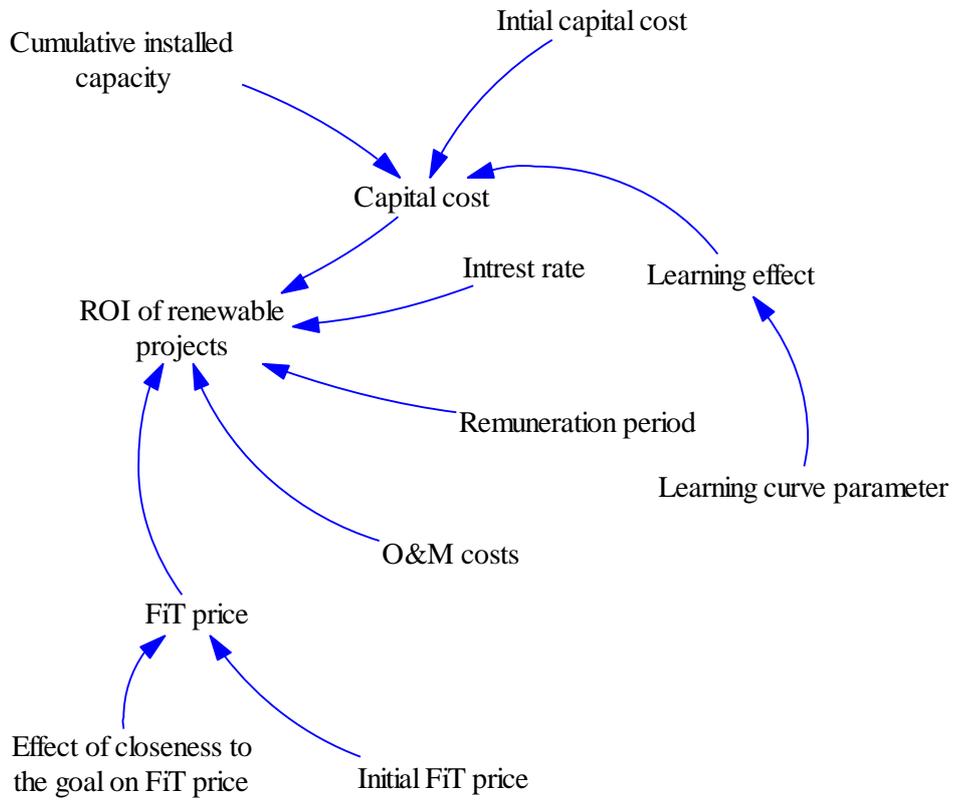

Fig. 6. ROI of renewable projects' causal relations.

### 5.3.2 FiT payment

Fig. 7 displays the stock-flow diagram of FiT payment subsystem. Installed capacity of REs multiplied by the capacity factor determines the electricity production in a year, which should be paid according to the FiT policy (see row 1 in Table 2). To calculate the money that should be paid by the government to electricity producers in each year, the total electricity production and the total FiT paid since the beginning of the simulation (2015) are accumulated in two stocks; the average FiT price is calculated by dividing the total electricity production by total FiT (see row 2 in Table 2). So, the average FiT price multiplied by electricity production determines the desired production payment for each year (see row 3 in Table 2). The initial values of both stocks are supposed to be zero at the beginning of the simulation.

Table 2. FiT payment subsystems' mathematical equations.

| Item | Variable (unit) | Mathematical equation |
|---|---|---|
| 1 | Electricity production (MWh/Year) | = Installed capacity * Capacity factor * 8760 |
| 2 | Average FiT price (Dollar/MWh) | = Total Electricity production/Total FiT payment |
| 3 | Desired production payment (Dollar/year) | = Electricity production * Average FiT price |

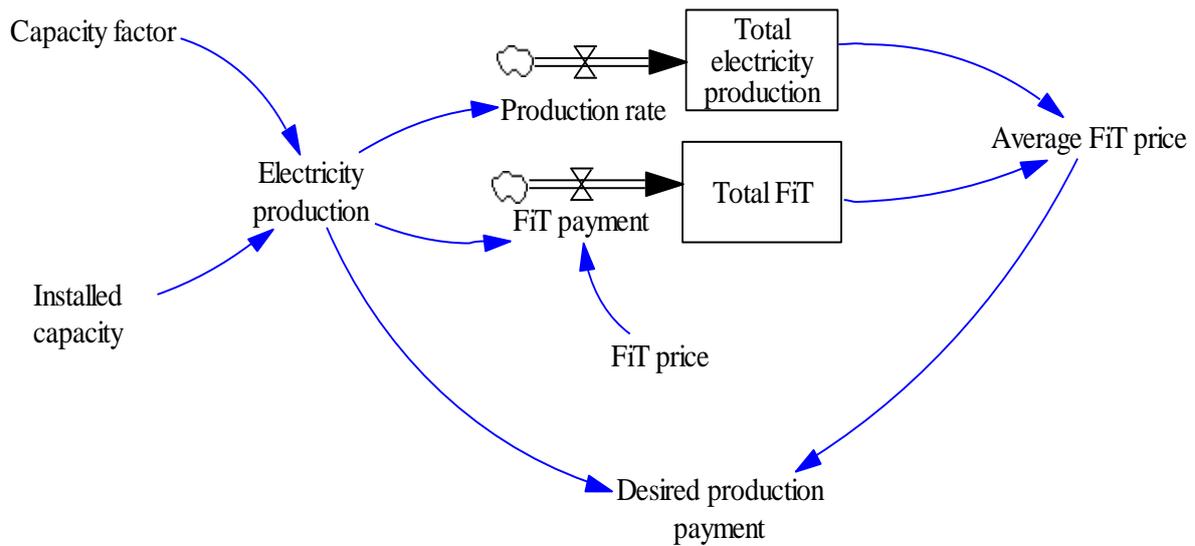

Fig. 7. Stock-flow diagram of FiT payment subsystem.

### 5.3.3 Budget

Fig. 8 displays the stock-flow diagram of the budget subsystem. The accumulated debt in the stock of SUNA debt plus the desired production payment, which is the output of FiT payment subsystem, determines the whole desired payment of the year (see row 1 in Table 3). If the whole desired payment is more than the amount of budget accumulated in the stock of budget, it would be possible to pay the whole desired payment; otherwise, all of the available budgets would be spent (see row 2 in Table 3). The available whole payment should be allocated to the production payment and debt payment with the priority of reducing the accumulated SUNA debt and then the production payment of the current year (see rows 3 and 4 in Table 3). SUNA debt is the cumulative amount of debt creation, which is rooted in the difference between the desired and actual production payments minus the debt payment (see rows 5, 6 and 7 in Table 3). Budget is the cumulative amount of budget increase minus budget decrease plus the initial value of the budget injected into the budget stock at the beginning of the policy implementation (see row 8 in Table 3). Budget decrease is defined as the summation of debt payment and actual production payment, and the budget increase is calculated by multiplying REs tax by electricity consumption (see rows 9 and 10 in Table 3). Electricity consumption is defined as an exogenous variable that is calculated by a linear regression equation through the simulation time horizon. The initial value of the budget is set as 2.5 million dollars [33]. Moreover, the initial value of SUNA debt equals to zero at the beginning of the simulation.

Table 3. Budget subsystems' mathematical equations.

| Item | Variable (unit) | Mathematical equation |
|---|---|---|
| 1 | Whole desired payment (Dollar) | = SUNA debt + Desired production payment |
| 2 | Available whole payment (Dollar) | IF THEN ELSE (Budget>=Whole desired payment, Whole desired payment, Budget) |
| 3 | Actual production payment (Dollar) | = IF THEN ELSE ((Available whole payment - SUNA debt) >= Desired production payment, Desired production payment, Available whole payment - SUNA debt) |
| 4 | Debt payment (Dollar/year) | IF THEN ELSE (Available whole payment>=SUNA debt, SUNA debt, Available whole payment) |
| 5 | Difference between the desired and actual production payments (Dollar) | = Desired production payment – Actual production payment |
| 6 | SUNA debt (Dollar) | = INTEG (Debt creation - Debt payment, 0) |
| 7 | Debt creation (Dollar/year) | = Difference between the desired and actual production payments |
| 8 | Budget (Dollar) | = INTEG (Budget increase - Budget decrease, 2.5e+06) |
| 9 | Budget decrease (Dollar/year) | = Debt payment + Actual production payment |
| 10 | Budget increase (Dollar/year) | = Electricity consumption (Time-based linear regression) * REs tax |

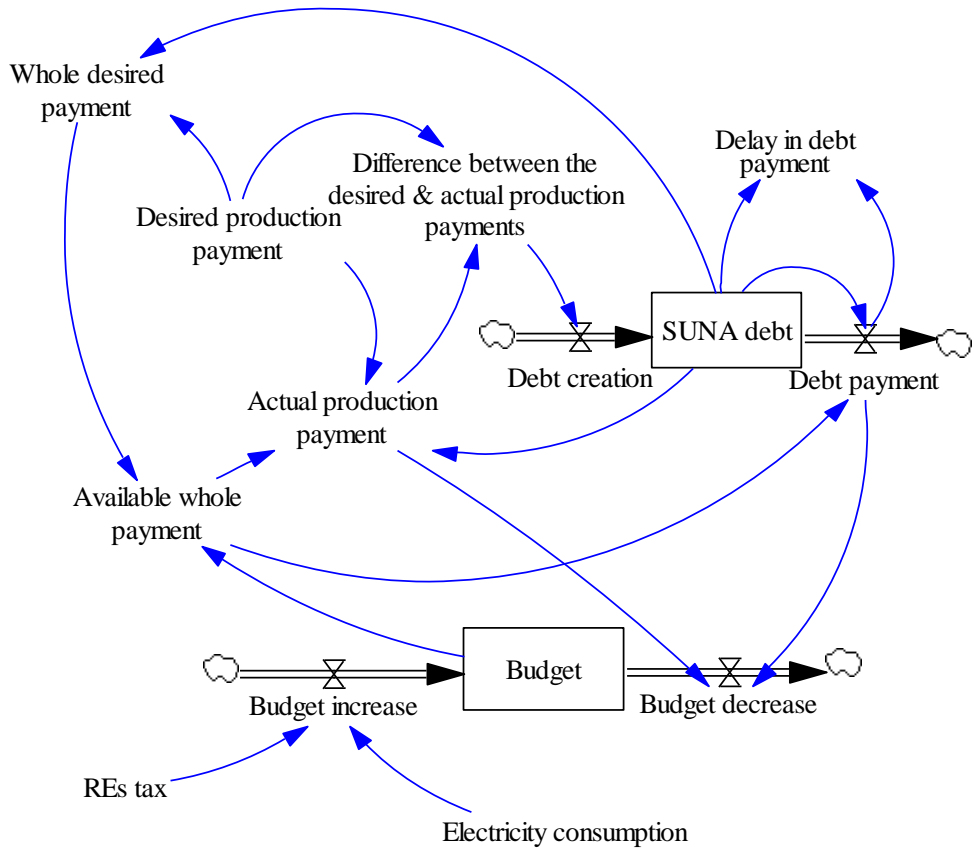

Fig. 8. Stock-flow diagram of the budget subsystem.

### 5.3.4 Social mechanisms

Some social effects are considered in the model that are rarely mentioned in the previous researches. These are the effect of delay in debt payment on the tendency to invest, and on O&M activities that the owner of the power plants do, and the effect of REs tax on social acceptance. Mathematically, all of these effects are formulated by an inverted sigmoid function with Y in the range of [0-1]. The mathematical expression of the function used in the model is depicted below:

$$Y = Y_{max}/[1 + (X/X_{50})^P] \tag{1}$$

where, Y is the value of the effect, $Y_{max}$ is the maximum value of the effect, X is the independent variable clarified for each specific effect, $X_{50}$ is X value at 50% value of Y, and P is an exponent to be found by optimizing and maximizing the model's goodness of fit to the existing data.

It is to be mentioned that there is no data for these social effects, and the numeric features of mathematical functions are determined based on SUNA experts' knowledge, energy policy researchers' viewpoints, and the content analysis of semi-structured interviews with renewable adopters. The visual form of this non-linear function is shown in Fig. 9. Conceptual details of each social mechanism and the numerical features of their effect mathematical functions are discussed below.

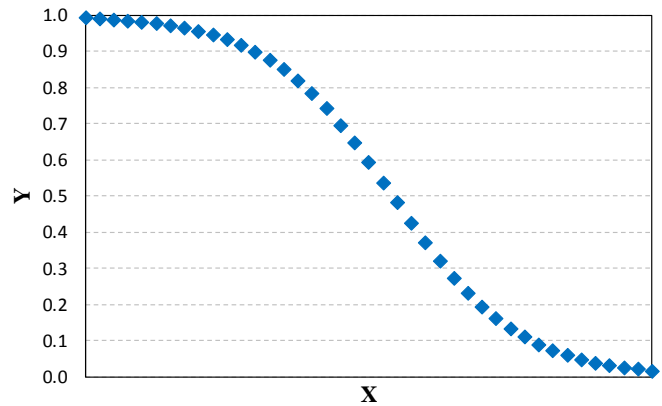

Fig. 9. Non-linear form of the inverted sigmoid function used in this research.

**Effect of REs tax on social acceptance:** When the government increases the REs tax, social acceptance decreases, which, indeed, represents the reaction of the investors to the amount of REs tax. We named this reaction "social tolerance". It is multiplied by the social acceptance value. When there is a low REs tax, its value is around one, representing no negative effect of low REs tax on social acceptance. On the other hand, in the extreme condition, when the REs tax increases to $0.1 per kilowatt hour (kWh) that is 100 fold of the current rate, a value near zero multiplied by social acceptance that reduces the social acceptance to near zero. It means that the policy makers could not increase the REs tax forever because the society has a tolerance threshold and is not neutral to REs tax rising. Clearly, other variables such as culture, education, and media might impact the social acceptance of REs by informing the people about their advantages like GHG mitigation effects, which are not considered here. The numerical features of this effect are shown in Table 4.

Table 4. Numerical features of social effects' functions.

| Y (Social effect) | X | $Y_{max}$ | $Y_{min}$ | X corresponds to $Y_{max}$ | X corresponds to $Y_{min}$ | $X_{50}$ | P |
|---|---|---|---|---|---|---|---|
| Effect of REs tax on social acceptance (Social tolerance) | REs tax ($/KWh) | 1 | 0 | 0 | 0.1 | 0.05 | 7 |
| Effect of delay in debt payment on tendency to invest (Potential investors' trust) | Delay in debt payment (Year) | 1 | 0 | 0 | 10 | 5 | 4 |
| Percentage of ideal O&M activities | Delay in debt payment (Year) | 1 | 0 | 0 | 10 | 5 | 6 |

**Effect of delay in debt payment on the tendency to invest**: When the accumulated debt of the government to RE producers increases, indeed, the delay in FiT payment increases so that the tendency of potential investors decreases. This concept is modeled by defining a variable named "potential investors' trust". It is assumed that when the debt delay in payment is close to 10 years, there would be no potential investors' trust, and also people's willingness to invest in new REs projects tends to zero.

**Effect of delay in debt payment on O&M activities:** When a producer is not paid on time, he/she may cut off some O&M activities in comparison with the ideal condition. We named this effect as "percentage of ideal O&M activities". While O&M activities decrease after a while, the equipment's lifetime decreases and depreciation rate rises causing more decline in the installed capacity. The numerical features of this effect are shown in Table 4.

## 5.4 Model validation

The validation process is critical for building confidence in a model's output. We followed validation methods and steps that the SD community subjects their models according to Qudrat-Ullah and Seong [34]. It is to be noted that both the structural and behavioral validity procedures are applied to the model.

### 5.4.1 Structural validation

**Boundary adequacy**

The model boundary adequacy was discussed in some meetings with the experts of SUNA and researchers in the field. Consistent with the purpose of the development of REs capacity, all the significant aggregates including installed capacity, budget, SUNA debt, annual FiT requests, approved FiT requests, capital cost of REs, ROI of renewable projects, tendency to invest, social acceptance, potential investors' trust, FiT price and electricity production from REs are generated endogenously. Total electricity generation capacity and electricity consumption are exogenous variables.

**Structure verification**

The structure verification of the model was tested by the available knowledge about the real system. The two knowledge source were SUNA data and experts' viewpoints.

**Dimensional consistency**

The dimensional consistency test requires testing all mathematical equations in the model and ensuring that the units of variables in each equation are consistent. We used "Unit Test" in Vensim and found that the model passed this test.

**Parameter verification**

The selection of parameter values determines the validity and feasibility of the model outcomes. The majority of values in this study are sourced from the existing knowledge and numerical data from SUNA. The remaining values are best guesses since no better data is available due to the fact that the policy implementation is on its infancy period. In addition, as the model is a high aggregate model, which addresses the REs development as a whole in the country of Iran, some parameters like normal equipment's lifetime, initial FiT price and the time needed to build are the average values of different REs types.

**Extreme condition test**

In this test, extreme values are assigned to the selected parameters, and then the model-generated behavior is compared to the reference (or anticipated) behavior of the real system under the same extreme conditions. We tested the model through two extreme-condition tests, and it was revealed that the outputs of the model were in line with the actual situation under extreme conditions, and its validity was enhanced.

Minimum governmental support: We set the remuneration period of the FiT policy as its minimum value that is 1 year, while the base value is 20 years. As an outcome, declining trend of installed capacity, no tendency to invest and a gradual growth of budget because of no payment for renewable electricity production were seen.

The financial burden at the beginning: It is supposed that a lot of debt (100 million dollars) exists at the beginning point of the policy implementation. There was an initial tendency to invest because of the policy announcement with good financial aspects; however, after the policy was started, it decreased to zero. Also, there was a steep slope for the budget because of the large payment for debt at the beginning.

**Structurally oriented behavior test**

Structurally oriented behavior or behavior sensitivity test was conducted and it was found that the fundamental patterns of behavior of the critical variables such as SUNA debt and installed capacity were insensitive to the parameters' change. Scenarios of increasing and decreasing the parameters, separately and also a mixture of increasing and decreasing them were carried out. The details of one of the scenarios are depicted in Table 5. The patterns generated by the model after these changes are shown in Fig. 10. The results indicated that changing the parameters could not alter the general behavior of the model. They could affect only some specific numerical values of the patterns such as a delayed take-off, or a higher peak.

Table 5. Parameters' change for structurally oriented behavior test.

| Parameter | Change |
| --- | --- |
| The time needed to build (year) | +70% |
| Normal equipment's lifetime (year) | +30% |
| Remuneration period (year) | +20% |

| Initial FiT price (Dollar/MWh) | -10% |
| Learning curve parameter (dimensionless) | -50% |

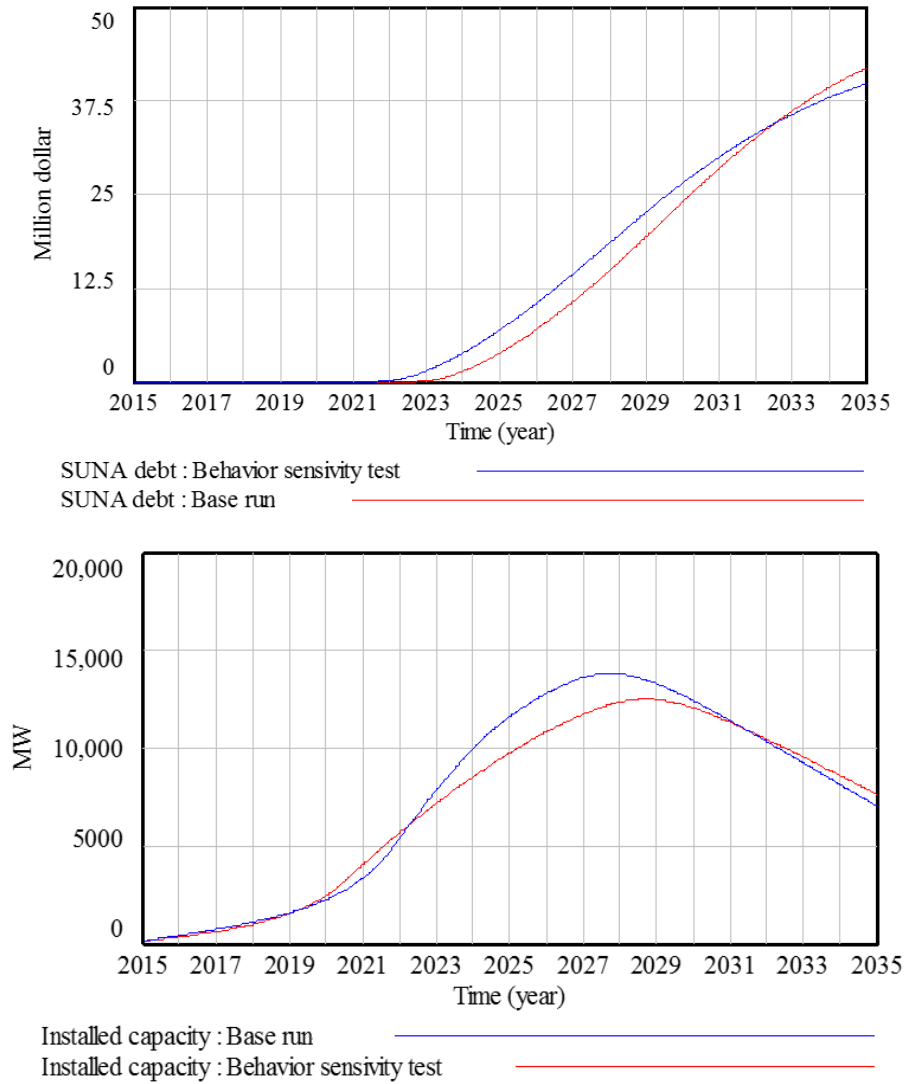

Fig. 10. Structurally oriented behavior test's behavior for SUNA debt and installed capacity.

### 5.4.2 Behavioral validity

The historical data are too narrow due to the fact that FiT policy has been implemented in Iran since 2015. Therefore, it is hard to find a reliable reference mode, and this model should be seen as a laboratory to do what-if analysis rather than a tool for accurate numeric forecasting. However, the two variables of "installed capacity" and "approved FiT requests" were selected to know how much the model could reproduce the historical data. As indicated in Figs. 11 and 12, the results of the simulation reproduce Iran's experience almost accurately regarding installed capacity, and approved FiT requests.

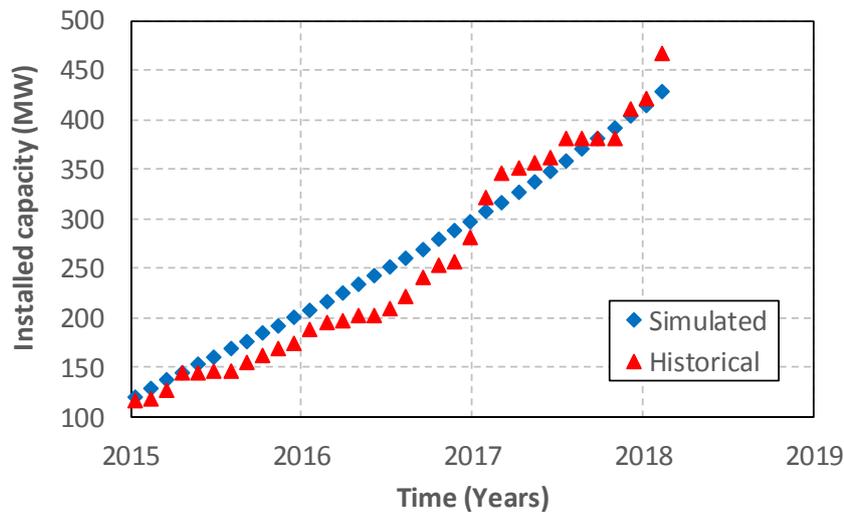

Fig. 11. Simulated and historical installed capacity.

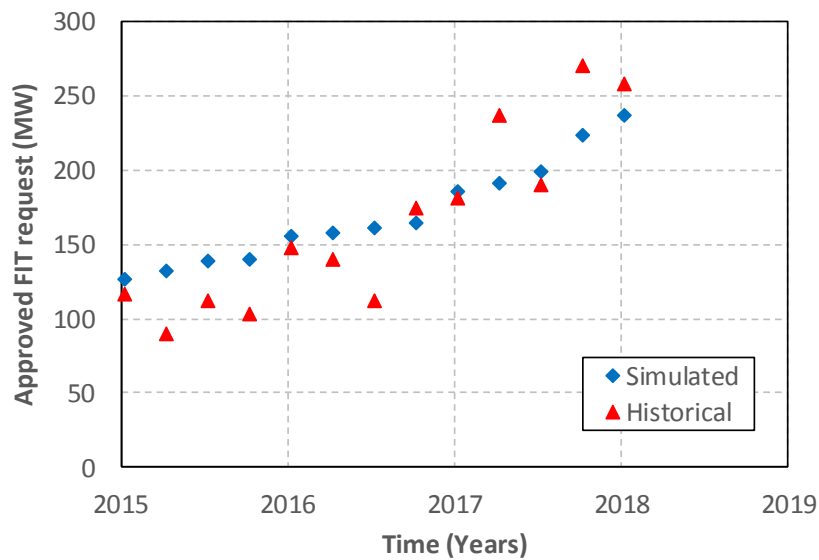

Fig. 12. Simulated and historical approved FiT requests.

The error analysis regarding the coefficient of determination ($R^2$), the mean squared error (MSE), the root mean squared percent error (RMSPE), and the Theil inequality statistics for these two variables is presented in Table 6. RMSPE provides a normalized measure of the magnitude of the error and MSE provides a measure of the total error. While the small total number of errors in the variables provides confidence in the model, large errors might suggest the presence of internal inconsistency of the model or the particular structure controlling the variables with significant errors. The Theil inequality statistics provide us with an excellent error decomposition to resolve such doubts [34].

Table 6. Error analysis of the model.

| Variable | $R^2$ | MSE (Units$^2$) | RMSPE (%) | $U^m$ | $U^s$ | $U^c$ |
|---|---|---|---|---|---|---|
| Installed capacity (MW) | 0.96 | 523 | 9 | 0.1 | 0.29 | 0.61 |
| Approved FiT requests | 0.89 | 891 | 23 | 0.11 | 0.7 | 0.19 |

$U^m$, $U^S$ and $U^C$ reflect the fraction of MSE due to bias, unequal variance, and unequal covariance, respectively.

Considering the installed capacity, $R^2$ is 0.96, showing a good ability of the model to reproduce the real historical data. RMSPE is 9%, which means that the variable replicates the behavior accurately. Of this small magnitude error, the significant portion (61%) is due to unequal co-variation, indicating that the simulated installed capacity tracks the underlying trend in the historical installed capacity almost perfectly but verges point-by-point. Considering the approved FiT requests, $R^2$ is 0.89, which shows a reliable behavioral reproduction ability of the model. Decomposition of the error statistics shows that the error is more rooted in unequal variation. According to Sterman [35], since the model's purpose is capturing the overall trend rather than the cycles and noises, the error could be unsystematic.

## 6 Simulation results

In this section, the simulation results of the model are analyzed. As mentioned before, the government's short-term target is reaching 5 GW in 2021, and the policymakers focus on this target rather than on long-term targets. Thus, through their short-term viewpoint, the simulation results are analyzed until 2021 and then long-term results are discussed. The target year (2021) is marked with a dashed line in the graphs.

### 6.1 Short-term future of REs

As shown in Fig. 13, the budget has an increasing trend up to 2020. Although its drop in the last year could be a sign of the system's altering state, SUNA debt is still zero, and financially, the system's performance is good, and the installed capacity will reach about 2,300 MW by the year 2021 (Fig. 14). Albeit it is less than half of the desired target, it has a favorable exponential trend and seems to reach the goal in the near future. The ROI of renewable projects, and consequently, the tendency to invest as the main stimuli for REs development declare a desired exponential growth trend of approximately 0.1 and 1.75, respectively (Figs. 15 and 16).

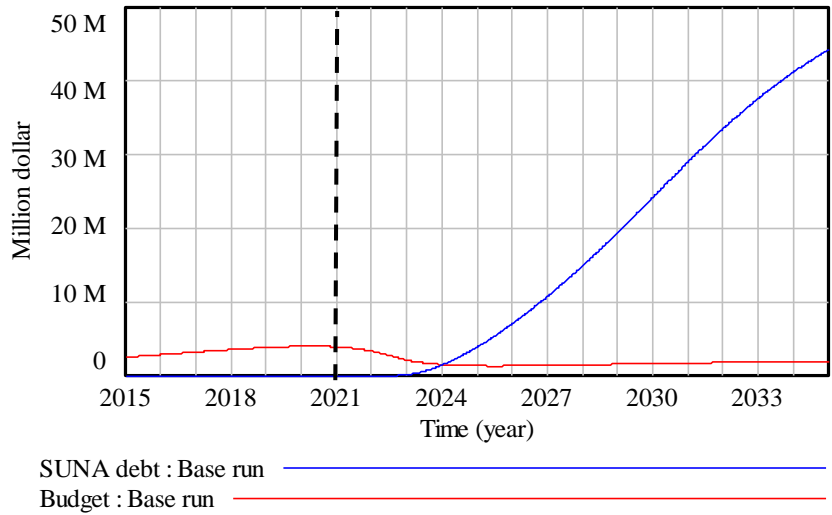

Fig. 13. Simulation results for SUNA debt versus budget.

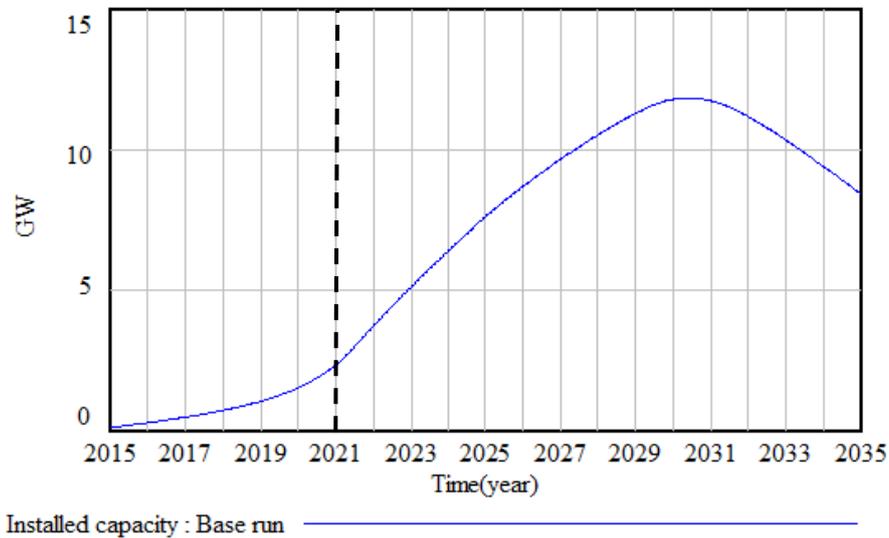

Fig. 14. Simulation results for installed capacity.

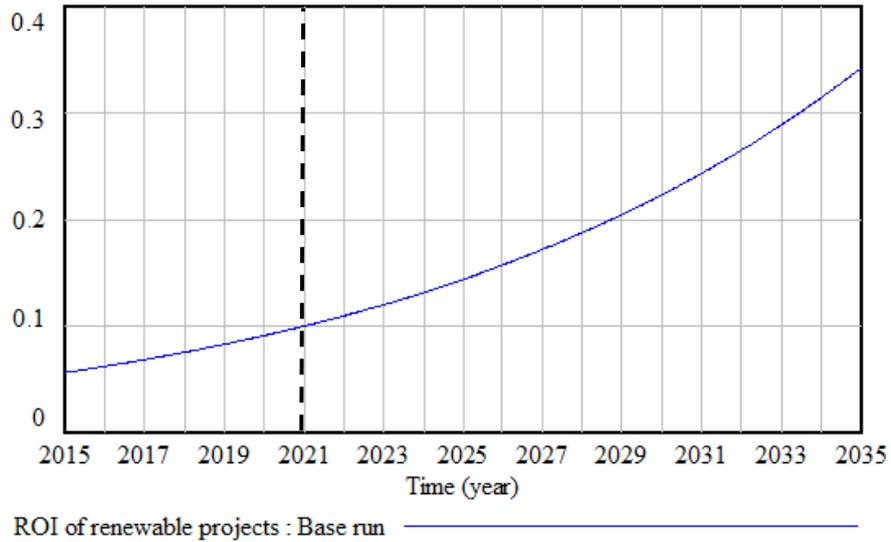

Fig. 15. Simulation results for ROI of renewable projects.

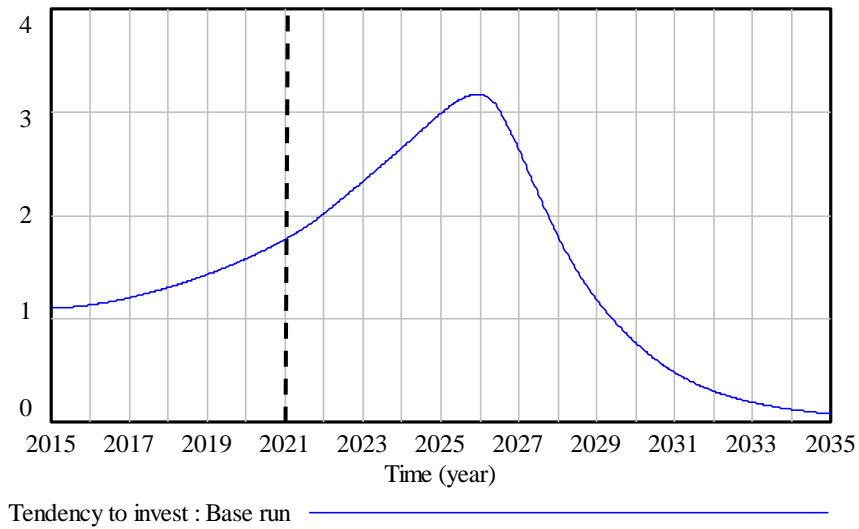

Fig. 16. Simulation results for the tendency to invest.

### 6.2 Expanding the time horizon

While everything looks desirable until 2021, expanding the time horizon to 2035 shows the different behavior of the system in long-term. SUNA debt rises since the year 2024, and the budget begins to reduce. In the year 2035, the difference between budget and debt would be about $40 million, meaning that the system will face a severe financial crisis (Fig. 13).

Only two years after the year 2021, the installed capacity will reach its desired target at 5 GW, and until then the exponential trend will remain unchanged, which may confuse the decision makers about the system's future behavior. After the year 2023, the behavior will gradually turn into an exponential decay. After reaching the peak of 12 GW at 2030, a dramatic decline will begin due to the depreciation rate overtaking the construction rate of installed capacity (Fig. 14).

Because of the social acceptance and learning reinforcing mechanisms, the ROI of renewable projects is on a significant rise. This variable is one of the important stimuli of the tendency to invest. Contrary to the expectations, the tendency to invest starts declining severely, due to the budget shortage and consequent SUNA debt increasing. The renewable producers sense this financial crisis through the delay in the time. They should be paid for the electricity they produced. This financial crisis triggers some social effects including reduced O&M activities by producers and a reduction in potential investors' trust, leading to the decline of a tendency to invest (Figs. 15 and 16).

## 7 Policy Analysis

In this section, the results from three policies considered for the FiT assessment model are discussed. The first policy is considered according to a short-term view of the issue, while the two other policies are based on a long-term view for sustainable development and taking the system feedbacks into account.

**Policy 1**: The first policy assumes a continuation of the current program without any structural change. Just the $0.03 increase in FiT price is considered in order to speed up the REs installed capacity development to achieve the desired goal at the target time (5 GW in 2021). It is a probable decision by the policymakers without a long-term systemic view.

**Policy 2**: In this policy, there would be a dynamic FiT price that is adjusted according to the budget status. It means that when there is a budget shortage in a specific year, FiT prices would be lowered, and when the government is financially wealthy, higher FiT prices would be announced.

**Policy 3**: Although SUNA believed that the amount of REs tax in the future would increase, due to the fact that in the year 2015 (which is the initial condition for this model), a considerable amount of budget was injected into the system, and apparently there was not a problem in the way of the REs development in the future, adjusting the budget based on the financial situation has not been considered seriously so far. A suggested policy to resolve the SUNA debt problem is getting feedback from the budget status to determine the amount of REs tax that is the entering rate of the budget stock. Policy 3 considers this issue.

Fig. 17 presents the amount of budget under policies 1, 2 and 3 compared to the base run. By applying Policy 1, the budget falls earlier compared with the base run. Higher FiT price causes a lower budget. The debt rises to $52 million that is approximately $6 million more than the base run; this means that it is in its worst-case. Regarding Policy 2, the amount of budget is considered to determine the FiT price. Hence, the budget falls smoother and later. However, after a while, the budget increases more steeply. In 2029, the SUNA debt will be about $1 million, which will be compensated by the budget in the next year and give a chance to the budget to rise again. Despite considering the budget status for determining FiT prices, there would be a little debt when Policy 2 is considered. The reason is that the budget shortage is perceived with delay, triggering the system to decrease FiT prices. When Policy 3 is applied, the increment amount of budget will be completely different from the former ones. While Policy 2 focuses on decreasing the debt, Policy 3 focuses on increasing the budget's input rate by rising REs tax rates. In this case, there would be no debt because the budget shortage would never happen.

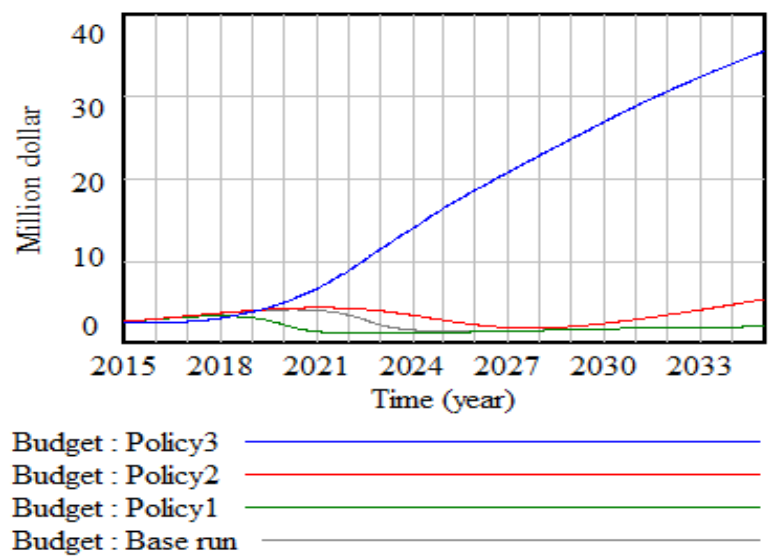

Fig. 17. Policy simulation results for the budget.

Fig. 18 presents the dynamics of the installed capacity growth under different policies. When Policy 1 is applied, the installed capacity reaches to 5 GW by the year 2021, which seems desirable for the policy-makers without a long-term view. This policy, sooner than all the other policies, makes the system fail, and the installed capacity faces a rapid drop after 2027. Considering Policy 2, although the installed capacity grows slower, taking feedback from the budget status, the rapid drop does not happen; instead, it follows a more stable trend. In addition, due to the budget increase that occurs in the year 2031, when the simulation duration increases, the stated drop is less. The installed capacity does not fall when Policy 3 is applied; it follows a favorable trend even with a later take-off.

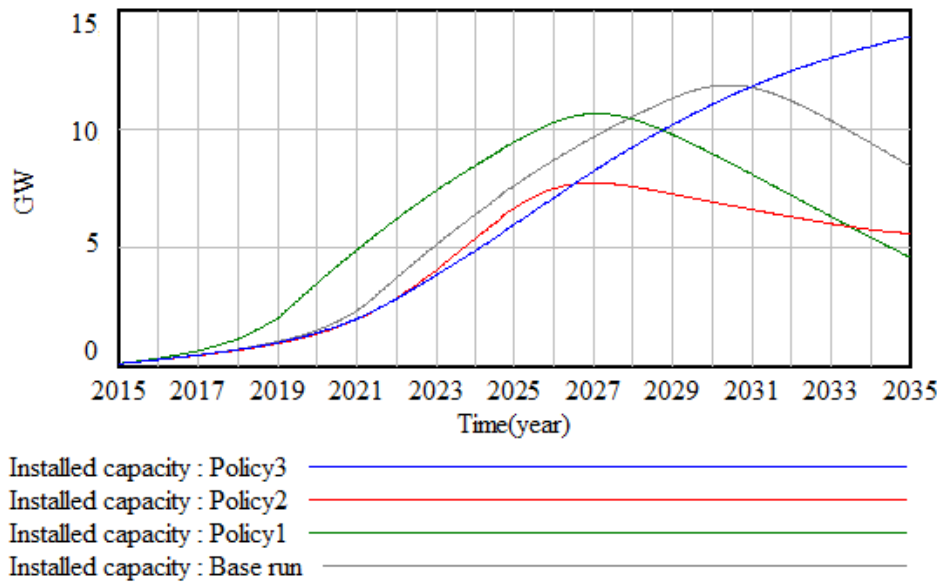

Fig. 18. Policy simulation results for installed capacity.

Figs. 19 and 20 present the tendency to invest and social acceptance under different policies. Regarding Policy 1, the tendency to invest is similar to the base run but the increase happens sooner, and finally, reaches nearly zero. The inefficiency of Policy 2 can be seen where the tendency to invest drops to near zero and then rises a little towards its value at the start of the simulation. As a consequence, there would be few FiT requests with Policy 2 implementation, implying that this policy can just avoid the budget shortage. By applying Policy 2 and by reducing FiT prices, the financial crisis will be prevented, but on the other hand, it means reducing the ROI of renewable projects, which causes investment attractiveness to fall, and therefore, lower tendency to invest. Policy 3 shows a favorable trend. Applying this policy, the tendency to invest increases up to 5 times by the year 2035. There is no debt to influence the tendency to invest negatively; hence the capital cost will decrease by the learning process, the decision makers will not be forced to reduce FiT prices, the ROI of renewable projects will increase and accordingly, the REs capacity will grow with a stable desirable trend.

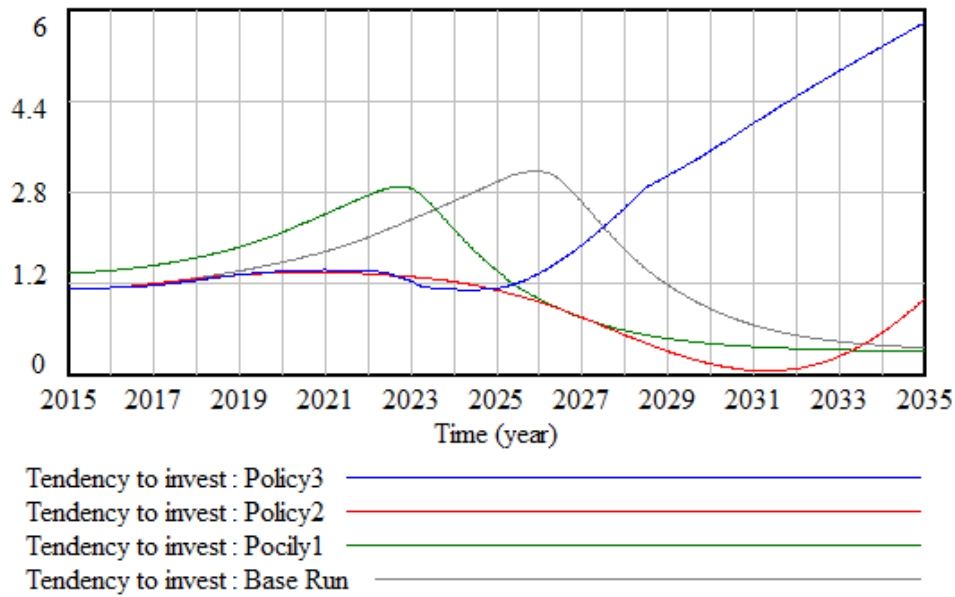

Fig. 19. Policy simulation results for the tendency to invest.

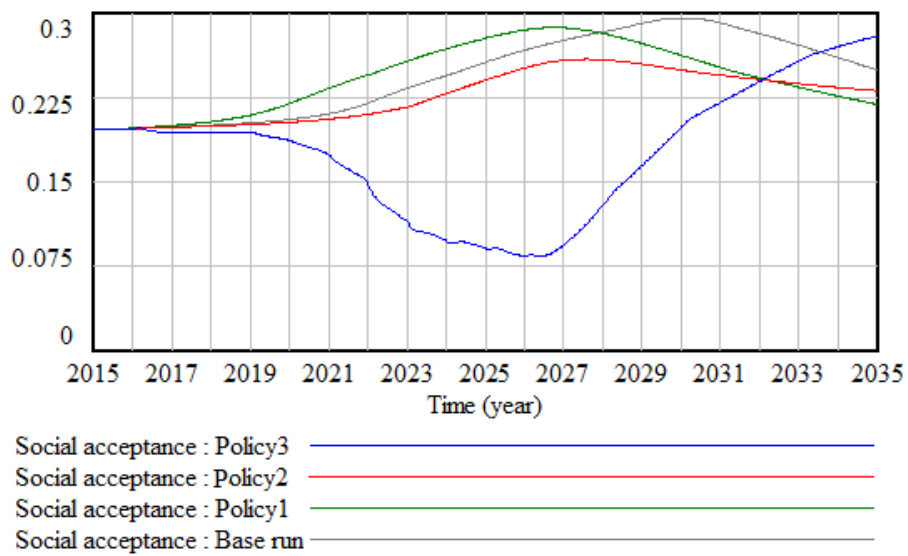

Fig. 20. Policy simulation results for social acceptance.

The reason why the tendency to invest starts to increase with a delay is rooted in REs tax rising and falling of social acceptance in the early years. When enough budget is funded for installed capacity development, the amount of REs tax gradually reduces, and social acceptance begins to rise, leading to more tendency to invest.

According to the policy analysis results depicted in Table 7, Policy 3 is the best one as prevents debt for SUNA, and therefore, avoids social effects caused by the debt. Moreover, it assures satisfying development of REs, reaching up to the amount of 14019.4 MW by the year 2035. Renewables' penetration rate reaches 0.13 meaning that 13% of the energy supply would be based on renewable resources. For an oil-dependent developing country like Iran, increasing the share of renewables from zero to 13% would be very promising.

Table 7. Policy simulation results for the year 2035.

| Variable (unit) | Base run | Policy 1 | Policy 2 | Policy 3 |
| --- | --- | --- | --- | --- |
| Installed capacity (MW) | 8434.11 | 4594.39 | 6069.77 | 14019.4 |
| Renewables' penetration rate (range of [0, 1]) | 0.08 | 0.04 | 0.05 | 0.13 |
| Tendency to invest (dimensionless) | 0.07 | 0.01 | 0.97 | 5.7 |
| SUNA debt (dollar) | 44200000 | 54200000 | 4081 | 0 |
| Delay in debt payment (year) | 23.07 | 28.27 | 0.99 | 0 |

## 8 Conclusions

Air pollution, energy security and increasing GHGs emission are some of the critical energy-related challenges for most countries. Development of REs is one of the most effective solutions to deal with these challenges. Although the FiT supporting policy is one of the most widely used policies to develop REs, it can lead to some financial problems. In this paper, Iran was selected as the case to show how the financial crisis could happen and how could governments prevent this by revising the FiT policy structure. Despite efforts made in recent decades by the government of Iran, REs situation is not desirable yet. Therefore, in 2015, the government implemented a FiT policy to develop REs and determined a target of 5 GW renewable installed capacity until 2021. This paper established an SD model to inquire whether this new FiT policy could assure the long-term growth of REs in Iran or it is just a temporary solution. By considering the social reactions to economic mechanisms and financial conditions in the system, we could make the model closer to the real world. The mentioned social reactions were the feedback of social acceptance to REs tax, the effect of the government's delay in payment on the tendency to invest and doing O&M activities.

The simulation results showed that the current program could not guarantee REs expansion in the long-term because of some malfunctions in the policy's financial structure. Three possible policies of 1) continuation of the current policy structure with a higher FiT price, 2) adjusting FiT price based on the budget status, and 3) adjusting REs tax upon the budget status were analyzed. The findings demonstrated that adjusting REs tax based on the budget status is the best policy among different policies. By applying this policy, the budget input rate increases with rising REs tax, there would not be any debt, the installed capacity will follow a favorable trend, social acceptance will rise after a while, and consequently, the system will follow an overall sustainable trend. The results can be generalized to any country that intends to implement FiT policies.

Future studies may consider the issue of developing competition between different types of REs. Mixing the proposed policies with probable scenarios will most probably widen the decision makers' perspective. Moreover, considering electricity demand and also the effect of increasing energy prices and taxes on electricity consumption as an endogenous mechanism can make the model closed to the real world.

**Acknowledgment**

This research was supported by the Renewable Organization of Iran (SUNA). The authors would like to thank Ms. Merla Kubli for her useful comments and suggestions to improve the quality of the paper.